\documentclass[12pt]{article}
\usepackage{graphics,color}
\begin{document}
\thispagestyle{empty}
\noindent\
\\
\\
\\
\begin{center}
\large \bf  Composite Weak Bosons at the LHC
\end{center}
\hfill
 \vspace*{1cm}
\noindent
\begin{center}
{\bf Harald Fritzsch}\\
Department f\"ur Physik\\ 
Ludwig-Maximilians-Universit\"at\\
M\"unchen, Germany \\
and\\
Institute of Advanced Studies\\
Nanyang Technological University\\
Singapore\\
\vspace*{0.5cm}
\end{center}

\begin{abstract}

In a composite model of the weak bosons  
the p-wave bosons are studied.
The state with the lowest mass is identified with the 
boson, which might have been detected recently at 
the LHC. Specific properties of 
these excited bosons 
are studied, in particular their decays into weak bosons, 
into photons and into leptons or quarks. 

\end{abstract}

\newpage

In the Standard Model of the electroweak interactions the masses of 
the weak bosons and of the leptons and quarks 
are generated by a spontaneous breaking of the electroweak symmetry. 
Besides the weak bosons a scalar boson must exist ( "Higgs boson").
Recently one has observed effects at the LHC, which might be due to the production of the
Higgs boson and subsequent decay into two weak bosons or into two photons (ref.(1,2)). 
The mass of this particle is about 126 GeV.\\

In this letter we assume that the weak bosons are composite particles. They consist 
of a lefthanded fermion and its antiparticle, 
which are denoted as "haplons".  A theory of this type was proposed in 1981 (see  ref.(3), also ref.(4,5,6,7,8)). 
The new confining chiral gauge theory is denoted as $QHD$. The $QHD$ mass scale is given by a mass parameter $\Lambda_h$, which determines the size of the weak bosons. The haplons interact with each other through the exchange of massless gauge bosons, which we denote as "glutinos" (the Latin expression 
for "glue" is "gluten"). \\

Two types of haplons are needed as constituents of the weak bosons, denoted by $\alpha$ and $\beta$.  
Their electric charges in units of e are:\\

\begin{equation}
h = \left( \begin{array}{l}
+\frac{1}{2}\\
-\frac{1}{2}\\
\end{array} \right) \ .
\end{equation}\\

The three weak bosons have the following internal structure:\\
\begin{eqnarray}
W^+ & = & \overline{\beta} \alpha \; , \nonumber \\
W^- & = & \overline{\alpha} \beta \; , \nonumber \\
W^3 & = & \frac{1}{\sqrt{2}} \left( \overline{\alpha} \alpha -
\overline{\beta} \beta \right) \; .
\end{eqnarray}\\
In the absence of electromagnetism the weak bosons are degenerate in mass. If the electromagnetic interaction is introduced,
the mass of the neutral boson increases due to the mixing with the photon. Details can be found in ref. (9,10).\\

The  $QHD$ mass scale is about thousand times larger than the  $QCD$ mass scale (ref.(9,10)). In strong interaction physics 
above the energy of 1 GeV many resonances exist. We expect similar effects in the electroweak sector. At high energies there 
should in particular exist excited weak bosons, which will decay mainly into two or three weak bosons. These states could 
be observed soon at the $LHC$.\\

The weak bosons consist of pairs of haplons, which are in an s-wave. The spins of the two 
haplons are aligned, as the spins of the quarks in a $\rho$-meson. The first excited states are those, in 
which the two haplons are in a p-wave. We describe the quantum numbers of these states by $I(J)$. 
The $SU(2)$-representation is denoted by $I$, $J$ describes the total angular momentum. There are three $SU(2)$ singlets , which we denote by S:\\

\begin{eqnarray}
S(0)=[0(0)], \nonumber \\
S(1)=[0(1)], \nonumber \\
S(2)=[0(2)]. 
\end{eqnarray}\\

The three $SU(2)$ triplets are denoted by T:\\

\begin{eqnarray}
 T(0)=[1(0)],  \nonumber \\
 T(1)=[1(1)], \nonumber \\
 T(2)=[1(2)].
\end{eqnarray}\\
\\
The internal structure of the $S$-states is given by:\\

\begin{eqnarray}
 S=& \frac{1}{\sqrt{2}} \left( \overline{\alpha} \alpha +
\overline{\beta} \beta \right) \ .
 \end{eqnarray}\\
\\
The three triplet states $T$ have the internal structure:\\
\begin{eqnarray}
T^+ & = & \overline{\beta} \alpha \; , \nonumber \\
T^- & = & \overline{\alpha} \beta \; , \nonumber \\
T^3 & = & \frac{1}{\sqrt{2}} \left( \overline{\alpha} \alpha -
\overline{\beta} \beta \right) \; .
\end{eqnarray}\\

We compare these states with the low mass mesons in strong interaction physics, in which the quarks 
are in a p-wave. We start with the scalar $\sigma$-meson (mass  $ \sim 600  -  800 $ MeV). This meson can easily mix with glue mesons, i.e. the 
$\sigma$-meson would be a superposition of a quark-antiquark-meson and a glue meson. 
This explains the relatively low mass of this meson.\\

The $\sigma$-meson is the QCD analogue of the state $S(0)$ - the spin one meson with isospin zero is $h_1(1170)$, analogous to the state S(1). The corresponding spin two meson is $f_2(1270)$, which is analogous to the state S(2):\\

\begin{eqnarray}
 S(0) \sim \sigma, \nonumber \\
 S(1) \sim h_1, \nonumber \\
 S(2) \sim f_2.
\end{eqnarray}\\

We expect that the scalar state $S(0)$ with the quantum numbers $[0(0^{+})]$ mixes with glutino bosons, analogous to the mixing of the $\sigma$-meson with glue mesons. Due to this mixing 
the mass of $S(0)$ is much lower than the masses of the other p-wave states.\\

Now we consider the isospin triplet mesons, the scalar meson $a_0(980)$, the vector meson $b_1(1235)$ and the spin two meson $a_2(1320)$. These mesons are analogous to the bosons T(0), T(1) and T(2) respectively:\\

\begin{eqnarray}
 T(0) \sim a_0, \nonumber \\
 T(1) \sim b_1, \nonumber \\
 T(2) \sim a_2.
\end{eqnarray}\\
\\
What is the mass spectrum of the excited weak bosons, in particular of the $S$ and $T$ bosons? The boson $S(0)$ should have a relatively small mass, as discussed above. We identify this 
boson with the particle, which might have been observed at CERN (ref. (1,2)). If the effect is real, the mass of $S(0)$ would be about 126 GeV. In what follows we shall assume: \\ 

\begin{equation}
M(S(0))= 126~GeV.
\end{equation}\\

In analogy to QCD we expect that the masses of the other p-wave states are in the range 0.2 - 0.6 TeV. The mixing of the bosons $S(1)$ and $S(2)$ with the glutino bosons is not as strong as the mixing for the $S(0)$. Thus we expect that the mass of the $S(1)$ - boson is just above 0.3 TeV, and the mass of the $S(2)$ - boson 
would be between 0.4 and 0.5 TeV.\\

The $SU(2)$ - triplet bosons $T$ cannot mix with glutino bosons. For this reason their masses should 
be larger than the masses of the $S$ - bosons. The mass of the $T(0)$ - boson should be about 0.3 TeV, 
the mass of the $T(1)$ - boson just above 0.4 TeV, and the mass of the  $T(0)$ - boson should be 
in the range 0.5  - 0.6 TeV. In the figure we show the mass spectrum of these bosons.\\

The $S(0)$ - boson will decay mainly into two charged weak bosons or into two $Z$-bosons (one 
of them virtual respectively) or into two photons. The $Z$-boson is the boson $W^3$, mixed with the photon. The mixing angle is 
the weak angle, measured to about 28.7 degrees. Using this angle, we can calculate the 
branching ratios BR for the various decays. The boson $W^3$ has a chance of 77 percent to decay into a $Z$-boson and 
a chance of 23 percent to decay into a photon. Thus we find the following branching ratios - the branching ratio for the decay into charged 
weak bosons is denoted by B: \\

$S(0) \Longrightarrow  ("W^+" + W^-) $    ~~~~~~~~~                BR  $\simeq B $, \\

$S(0) \Longrightarrow  ("W^-" + W^+) $     ~~~~~~~~~              BR  $\simeq B $, \\  

$S(0) \Longrightarrow  ("Z" + Z) $     ~~~~~~~~~              BR  $\simeq 0.59 ~B $, \\
 
$S(0) \Longrightarrow  ( Z +  \gamma ) $      ~~~~~~~~~~~~~~~~~~             BR $\simeq 0.36 ~B $, \\
 
$S(0) \Longrightarrow  (\gamma + \gamma) $    ~~~~~~~~~~~~~~~~~~~        Br$\simeq 0.05 ~B$ \\ 

- "Z" means  "virtual Z" etc. \\  
  
We would not expect that the decay rates for the decays of $S(0)$ into letons and 
quarks are given by the mass of the fermion, 
as they are for the Higgs boson. The branching ratio for the decay into an 
electron pair, into a muon pair, into a 
tau pair or into a neutrino pair would be the same: \\

BR ($S(0) \Longrightarrow  (e^+ + e^-) $ ) $\simeq $ BR ($S(0) \Longrightarrow  (\mu ^+ + \mu^-) $ ) \\

$\simeq $ BR ($S(0) \Longrightarrow  (\tau ^+ + \tau^-) $ ) $\simeq $ BR ($S(0) \Longrightarrow  (\nu + \bar{\nu}) $ ) \\

( $\nu$ stands for the electron, muon and tau neutrino ). \\

Due to the color of the quarks the branching ratio for the decays into quarks are three times larger:\\

BR ($S(0) \Longrightarrow  (u + \bar{u}) $ )  $\simeq $ 3 BR ($S(0) \Longrightarrow  (e^+ + e^-) $ ) \\

BR ($S(0) \Longrightarrow  (d + \bar{d}) $ )  $\simeq $ 3 BR ($S(0) \Longrightarrow  (e^+ + e^-) $ )\\

BR ($S(0) \Longrightarrow  (u + \bar{u}) $ ) $\simeq $ BR ($S(0) \Longrightarrow  (c + \bar{c}) $ )\\ 
    
BR ($S(0) \Longrightarrow  (d + \bar{d}) $ ) $\simeq $ BR ($S(0) \Longrightarrow  (s + \bar{s}) $ ) \\ 

$\simeq $ BR ($S(0) \Longrightarrow  (b + \bar{b}) $ ).\\    

Presumably the $S(0)$ decays mainly into two weak bosons or into two photons. As an example let us suppose that the branching ratio 
for the decay into weak bosons and photons is 70 \%. Then the branching ratios for the decays into leptons and quarks 
would be 30 \%. Since there are 21 different fermion channels, the branching ratio for the 
decay into a muon pair would be 1.4 \%.\\

We expect that decays of the  $S(0)$ - boson into a pair of muons should be observed soon at the LHC. The branching ratio for this decay should be about 30 \% of the branching ratio for the decay into two photons.\\

The bosons $S(1)$ and $S(2)$ have a much higher mass as the $S(0)$ - boson. They will decay 
mainly into three or four weak bosons, e.g.: \\

$S(1) \Longrightarrow  (W^+ + W^- + Z) $,\\
$S(1) \Longrightarrow  (W^+ + W^- + W^+ + W^-) $,\\
$S(1) \Longrightarrow  (W^+ + W^- + \gamma) $.\\

Decays into two weak bosons, two photons or into lepton pairs or quarks pairs would be suppressed.\\

The $SU(2)$ - triplet bosons $T(0)$, $T(1)$ and $T(2)$ will decay mainly into four or five weak bosons or photons. 
Decays into two weak bosons or two photons are strongly suppressed, also the decays into lepton or quark pairs.\\

The boson $T^+$ or $T^-$ can in principle decay into a charged weak boson and a $Z$-boson or a 
photon, however the branching ratio for this decay will be very small. The main decay modes of the 
$T^+$ would be:\\

$T^+ \Longrightarrow  (W^+ + Z + Z) $,\\

$T^+ \Longrightarrow  (W^+ + Z + \gamma ) $,\\

$T^+ \Longrightarrow  (W^+ + \gamma + \gamma) $.\\

The properties of $S(0)$ with a mass of 126 GeV, which might have been detected 
at the LHC recently,  should be investigated in detail. If the model, discussed here, has something to do with reality, the bosons $S(1)$ and $T(0)$ should be found during the next two years at the LHC.\\  

Acknowledgement: This paper was written during my visit in the Institute for Advanced Study 
at the Nanyang Technological University in Singapore. I would like to thank Prof. K. K. Phua for 
the support.

\end{document}